\documentclass[twocolumn]{revtex4}
\usepackage{graphicx} 
\usepackage{titlesec}
\titleformat*{\section}{\large\bfseries\sffamily}
\titleformat{\subsection}[runin]
  {\normalfont\normalsize\bfseries}{\thesubsection}{1em}{}
\titlespacing*{\section}
{0pt}{5.5ex plus 1ex minus .2ex}{0.3ex plus .2ex}
\usepackage{color}
\usepackage[normalem]{ulem}
\usepackage{soul,xcolor}
\setstcolor{red}

\begin{document}

\title{Zoology of domain walls in qausi-2D correlated charge density wave of 1T-TaS$_{2}$}

\def\correspondingautho{\footnote[1]{Corresponding author: absolute81@ibs.re.kr}}
\def\correspondingauthor{\footnote[2]{Corresponding author: yeom@postech.ac.kr}}
\author{Jae Whan Park$^1$ \correspondingautho{}}
\author{Jinwon Lee$^{1,2}$}
\author{Han Woong Yeom$^{1,2}$ \correspondingauthor{}}
\affiliation{$^1$Center for Artificial Low Dimensional Electronic Systems,  Institute for Basic Science (IBS),  77 Cheongam-Ro, Pohang 790-784, Korea.} 
\affiliation{$^2$Department of Physics, Pohang University of Science and Technology, Pohang 790-784, Korea}

\begin{abstract}

Domain walls in correlated charge density wave compounds such as 1T-TaS$_{2}$ can have distinct localized states which govern physical properties and functionalities of emerging quantum phases.
However, detailed atomic and electronic structures of domain walls have largely been elusive.
We identify using scanning tunneling microscope and density functional theory calculations the atomic and electronic structures for a plethora of discommensuration domain walls in 1T-TaS$_{2}$ quenched metastably with nanoscale domain wall networks.
The domain walls exhibit various in-gap states within the Mott gap but metallic states appear in only particular types of domain walls. 
A systematic understanding of the domain-wall electronic property requests not only the electron counting but also including various intertwined interactions such as structural relaxation, electron correlation, and charge transfer.
This work guides the domain wall engineering of the functionality in correlated van der Waals materials.  

\end{abstract}

\maketitle


Metal-insulator transitions and the emergence of superconductivity and other exotic quantum phases from strongly interacting electronic systems are central issues in condensed matter physics and important for novel device applications.
Such interacting electronic systems are recently attracting particular interest in two dimensional (2D) materials with the extraordinary tunability provided \cite{wood14,cao18,naik2018}. 
In particular, quasi 2D charge-density-wave (CDW) compounds and their monolayer versions with strong interactions are under focus of huge research activity because of rich quantum phases emerging from the commensurate CDW (CCDW) ground state with exotic electronic properties such as a spin frustrated Mott insulator in 1T-TaS$_{2}$ and an excitonic insulator in 1T-TiSe$_{2}$ \cite{faze80,may11}.
Suppressing the CCDW order in these materials by heat, pressure, doping, ultrafast optical and electrical pulses leads to various quasi-metallic metastable phases which consist of CCDW domains and discommensuration domain walls (DWs) \cite{yan17,joe14,stoj14, yosh15, tsen15, cho16, ma16, vask15, vask16, xu10, Ang12, bu19, han15,liu16,zong18,wen19,saka17,laul17,avig18,sipo08, li16, ang12,chen18}.
The electronic properties of these `frustrated' phases have been known to be governed by electronic states residing on DWs created \cite{liu16,sipo08, li16, ang12,chen18}.
Those topological in-gap states have been suggested to play crucial roles in the emerging superconductivity \cite{sipo08,li16,chen18,park19} and the memristic resistivity switching of DW phases  \cite{yosh15}.


In the present work, we focus on the DW phases excited from the Mott CDW ground state of 1T-TaS$_{2}$ \cite{pill00, cler06, cho15} with strong electron correlation \cite{faze80} and strong spin frustration \cite{law17, klan17}.
The atomic structure of CCDW in  1T-TaS$_{2}$ has unit of the David-star (DS) cluster with a $\sqrt{13}\times\sqrt{13}$ supercell \cite{brou80}, which features one unpaired electron in the central Ta atom. 
This electron has been known to fall into a spin-frustrated Mott-insulator state by the onsite Coulomb repulsion \cite{faze80}.
Recent STS study observed two types of isolated DWs among the various types of DW configuration \cite{cho16,ma16}, as line defects within this `correlated' CCDW phase, which have insulating electronic states due to strong electron correlation and structural relaxation within them \cite{cho17}.
Above 180 K, the Mott insulating CCDW phase transits spontaneously into a metallic phase with a nanoscale honeycomb DW network, which is called as a nearly-commensurate CDW (NCCDW) phase \cite{wu89, naka77, yama83, wu90, burk91, spij97, park19}. 
Our recent work identified this DW with another type of structures and a metallic in-gap state \cite{park19}, which is absolutely different from two types of insulating DWs within the CCDW phase \cite{cho17}. 
A theoretical model was suggested \cite{park19,lee20} to connect it to the emerging superconductivity from this phase under pressure and doping \cite{sipo08}.
These series of recent research revealed the various electronic properties and huge structural degrees of freedom within DWs of 1T-TaS$_{2}$ CDW phases. 
However, only a small fraction of DWs are identified among those expected from the large (at least thirteen-fold) degeneracy of the CDW structure and a systematic understanding of their diversity and complexity has been lacking. This information would be essential for the controllability over the functionality and the emerging or hidden phases brought by DWs \cite{stoj14,vask15,vask16,han15}.  


In this paper, we explore structural and electronic properties of the 1T-TaS$_{2}$ surface layer, which was manipulated by scanning tunneling microscope (STM) voltage pulses into nanoscale metastable DW networks.
The metastable DW networks are endowed with an enhanced variety of DWs \cite{cho16,ma16}. 
We identify more than ten different DW structures and their topological vertices. 
The atomic structures and electronic states of most of these DWs are determined by direct comparisons of STM data and density functional theory (DFT) calculations.
A systematic understanding of the structural and electronic variety of DWs are obtained through extensive DFT calculations, which unravel the diverse interactions involved such as structural relaxations, electron correlation effect, and charge transfer from neighboring domains. 


\begin{figure*}[t]
\centering{ \includegraphics[width=18.0 cm]{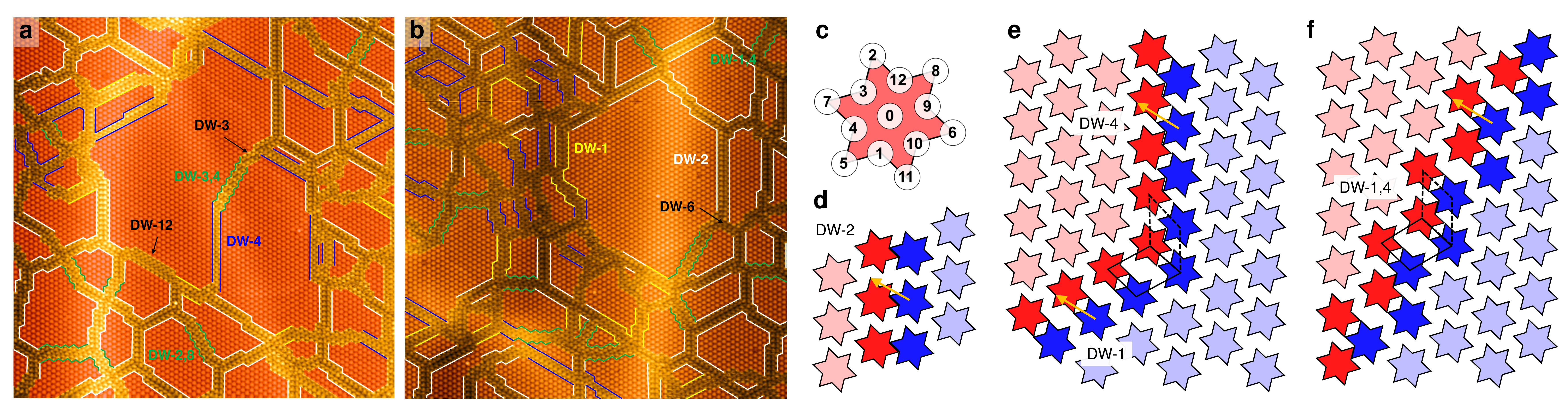} }
\caption{ \label{fig1}
Domain and domain-wall structures in quenched 1T-TaS$_{2}$.
{\bfseries\sffamily a} and {\bfseries\sffamily b} Two independent STM images (sample bias voltage $V_s$ = -1.2 V, tunneling current $I_t$ = 10 pA, and size $L^2$ = 80 $\times$ 80 nm$^2$).
The line colors denote the type of domain wall: DW-1: yellow, DW-2: white, DW-4: blue, and $zigzag$ domain wall: green.
{\bfseries\sffamily c} David-star cluster of the CCDW phase with thirteen Ta atoms indexed. A DW is defined as the center of DS clusters of the neighboring domain sits on Ta atomic sites of 1-12, which is indexed as DW-1-DW-12, respectively.  
{\bfseries\sffamily d}-{\bfseries\sffamily f} Schematic arrangement of David stars for the domain walls.
{\bfseries\sffamily d} DW-2,
{\bfseries\sffamily e} DW-1 and DW-4, and
{\bfseries\sffamily f} $zigzag$ domain wall (DW-1,4).
Orange arrows denote the CCDW vector.
Solid (dashed) parallelogram in {\bfseries\sffamily e} denotes the domain-wall unit of DW-1 (DW-4).
}
\end{figure*} 

\section*{Results and Discussion}
\subsection*{Domain walls and their network in the nanoscale quenched phase.}
Various types of DWs are found in the quenched phase of Fig. 1a and 1b, which are created by applying a positive voltage pulse from a STM tip \cite{cho15}.
It consists of domains of a few tens of nanometers in the CCDW structure as separated by various different DWs.
These DWs form an irregular network extending laterally to a size of up to a few hundred nanometers depending on the pulse voltage. 
Most DWs are formed straightly along the CCDW unit vector with a single phase shift between two adjacent CCDW domains.
Due to the thirteen Ta atoms in the DS cluster, there are twelve possible $straight$ DW configurations and the notation of DWs are defined as the relative position of the center DS clusters between two adjacent domains as shown in Fig. 1c and Supplementary Fig. 1.
We also found the $zigzag$ type of DWs, which can be defined by two alternating phase-shift vectors, one of which is rotated by 60$^{\circ}$  with respected to the other (see Fig. 1f).
Thus, the $zigzag$ DWs are rotated by 30 $^{\circ}$ with respect to the $straight$ DW.
The twelve possible $zigzag$ DW configurations are summarized in Supplementary Fig. 1.
The DWs also construct the complex DW junctions including the X, Y, and +  shape (Supplementary Fig. 2).
The honeycomb-like network of DWs can be seen as the combination of Y-shaped junctions (Supplementary Fig. 3).
The detailed properties of DWs will be discussed further below.

\subsection*{Population and energetics of domain walls.}
Among the twelve possible $straight$ DW configurations, we found six types of DWs in the quenched phase.
The second-type DW (DW-2) is the most popular DW structure and DW-4 and DW-1 are also frequently found.
These DWs were already observed in the CCDW phase at low temperature (DW-2 and DW-4) and in the NCCDW phase at room temperature (DW-1) \cite{cho16,cho17,park19}.
Three minor DWs of DW-3, DW-6 and DW-12 appear sparsely with shorter lengths, but other six possible $straight$ DWs cannot be found in the present STM images.
Interestingly, the $zigzag$ type of DWs are rather popular (less than DW-4 but more than DW-1) with relatively long DW lengths.
We found three types of $zigzag$ DWs, namely, DW-2,8, DW-1,4 and DW-3,4. 
The two indices given here indicate the two phase shift vectors defining a $zigzag$ DW (Fig. 1f).
We optimize atomic structures of all twelve configurations of $straight$ DWs by DFT calculations.
The resulting energetics and electronic properties are summarized in Table 1.
The fully relaxed DFT calculations predict that DW-2 is the most stable structure and the second-lowest energy configurations of DW-1 and DW-4 are almost degenerated in energy.
This energetic hierarchy is largely consistent with the observed population of major DWs. 
However, the present calculations cannot explain all the population details of the minor DWs of the DW-3, DW-6, and DW-12 and the other absent DWs in the experiment. 
This may be due to their short lengths and various boundary conditions imposed by neighboring domains, DWs and DW junctions.
In addition, we also observe that the energetically unfavored DWs, such as DW-5, DW-8, and DW-6 break into two DWs of lower energies as shown in Fig. 2. 
This partly explains the absence of the unfavored DWs.
The energy costs of $zigzag$ DWs (0.025, 0.028, and 0.030 eV/DS for DW-2,8, DW-1,4, and DW-3,4, respectively) is comparable to those of major straight DWs.
This seems roughly consistent with their popularity in the experiment, being popular than the minor straight DWs. The present results indicate the importance of detailed atomic structures in understanding the DW phases. 


\begin{figure*}[t] 
\centering{ \includegraphics[width=16.0 cm]{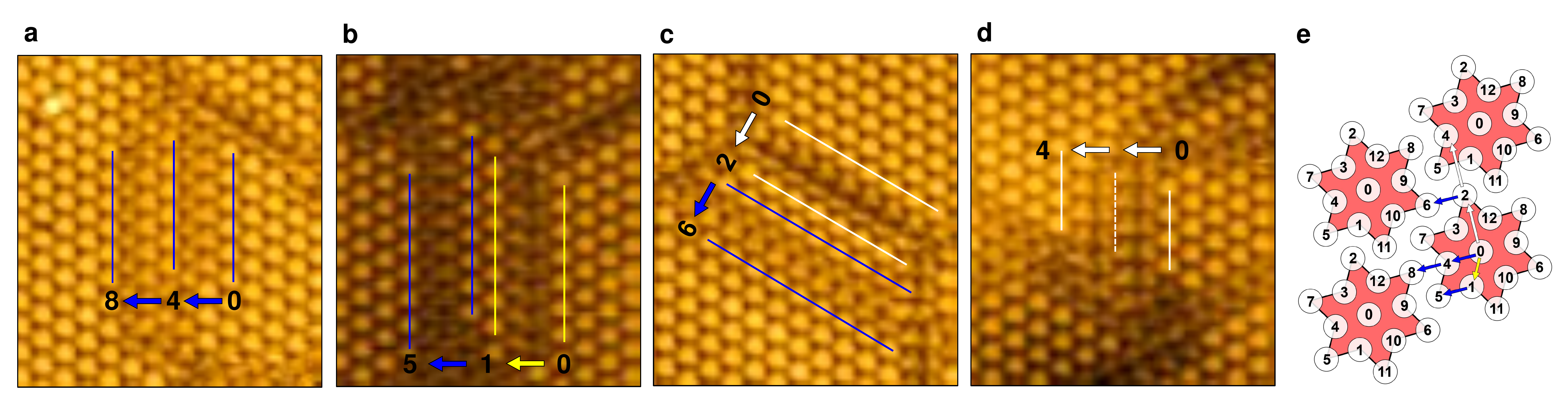} }
\caption{ \label{fig2}
Narrow domains between stable domain walls.
{\bfseries\sffamily a} A single-row domain with two DW-4 structures instead of DW-8.
{\bfseries\sffamily b} A double-row domain with DW-1 and DW-4 structures instead of DW-5.
{\bfseries\sffamily c} A double-row domain with DW-2 and DW-4 structures instead of DW-6.
{\bfseries\sffamily d} Direct connected two DW-2 structures without domain.
{\bfseries\sffamily e} Corresponding phase shift vectors on the David-star clusters.
}
\end{figure*} 

\subsection*{Atomic structures of straight domain walls.}
The atomic structure of DW-2 in Fig. 3b is composed of two imperfect DS clusters of a similar shape, each with 12 Ta atoms but with distinct numbers, five (left) and six (right), of S atoms.
Since bright protrusions in STM images represent top-layer S atoms, the different numbers of S atoms are well reflected in the corresponding STM images; larger triangular clusters and a smaller imperfect triangles in the left and the right side of the DW, respectively (Fig. 3f).
These STM images are well reproduced in the simulated STM images, especially the asymmetric double rows along the DW.
The present double row structure is distinguishable from the previously suggested model with a single row \cite{cho17}.
These two models are very close in their energies, only 9 meV difference per supercell, and their electronic structures are very similar.
Thus, these structures may coexist (Supplementary Fig. 4), while we find the double row model to reproduce better the STM images observed.

The STM image of DW-4 exhibits also an asymmetric double row (named as DW-4B) of imperfect DS clusters with 22 Ta atoms (the lower part of Fig. 3g).
Very interestingly, we find that this double row structure changes abruptly to a single row structure (DW-4A) with the same phase shift between neighboring domains (Fig. 3g).
This evidences that a DW can have a few competing structures, due to its own internal structural degree of freedom.
Our DFT calculation tells that the single-row structure (DW-4A) is formed with a single imperfect DS cluster (with nine Ta atoms) and distorted DS clusters at the neighboring domain edge.
This structure is consistent with the previous structural model \cite{cho17}, and it is more stable ($\sim$ 56 meV/supercell) than the double row (DW-4B) structure.
The optimized atomic structures of DW-4A and DW-4B reproduce the observed STM images well. We also find another type of a single-row DW structure without a strong distortion of the domain edge, which is slightly less stable ($\sim$ 29 meV/supercell) than the above DW-4A structure.
However, its electronic structure is not consistent with previous STS spectra \cite{cho17} (Supplementary Fig. 9).

The DW-1 structure comprising 25 Ta atoms is similar to the DW-2 but features an additional Ta atom between two imperfect DS clusters.
The corresponding STM image shows rows of three clusters within DW-1 in contrast to the double rows of DW-2 (Fig. 3e).
This structure is consistent with our previous work \cite{park19}. 
Other minor DWs also have their characteristic atomic structure, which are determined by their own CDW phase shift vectors and the internal structural relaxation (Supplementary Fig. 5 and 6). 

\begin{table*}[t]
\caption{ \label{table1}
Energetics and electronic structure of all twelve $straight$ domain walls.
Formation energy is defined by  $E$ = $E$$_{DW}$/$N$ $-$ $E$$_{CCDW}$,  where $E$$_{DW}$ and $E$$_{CCDW}$ are the total energies and $N$ is the number of the David-star clusters in the supercell.
$E$$_{PBE}$ and $E$$_{PBE+U}$ are obtained from the exclusion and inclusion of the extra electron correlation U, respectively.
LHS and UHS are the lower and upper Hubbard states at domain region of the supercell, respectively. 
For the CCDW phase, the LHS (UHS) locates at -0.216 (+0.198) eV.
M and I stand for metal and insulator.
}
 \begin{tabular}{c| c c c c c c c c c c c c }\hline\hline
  & DW-7 &  DW-5 & DW-4A (DW-4B)  & DW-3 & DW-2 & DW-1 &  DW-12 & DW-11 & DW-10 & DW-9 & DW-8 & DW-6\\\hline
Number of Ta  & -7 & -5 & -4 & -3 & -2 & -1 & 1 & 2 & 3 & 4 & 5 & 7\\
E$_{PBE+U}$ (eV/DS)  & 0.107 &  0.085 & 0.033 (0.044) & 0.058 & 0.025 & 0.039 & 0.069 & 0.059 & 0.042 & 0.058 & 0.047 & 0.048\\
E$_{PBE}$ (eV/DS)  & 0.079 &  0.069 & 0.022 (0.027) & 0.040 & 0.010 & 0.023 & 0.061 & 0.045 & 0.052 & 0.063 & 0.038 & 0.068\\
LHS & -0.080 &  -0.057 & -0.042 (-0.054) & -0.126 & -0.121 &-0.041 & -0.107 & -0.226 & -0.087 & -0.172 & -0.098 & -0.084 \\
UHS & 0.301 & 0.302 & 0.320 (0.322) & 0.270 & 0.251 & 0.327 & 0.270 & 0.144 & 0.309 & 0.219 & 0.274 & 0.290 \\
character  & M & M & I (M) & I & I & M & M & M & I & I & I & I \\\hline\hline
 \end{tabular}
\end{table*}


\begin{figure*}[t] 
\centering{ \includegraphics[width=18.0 cm]{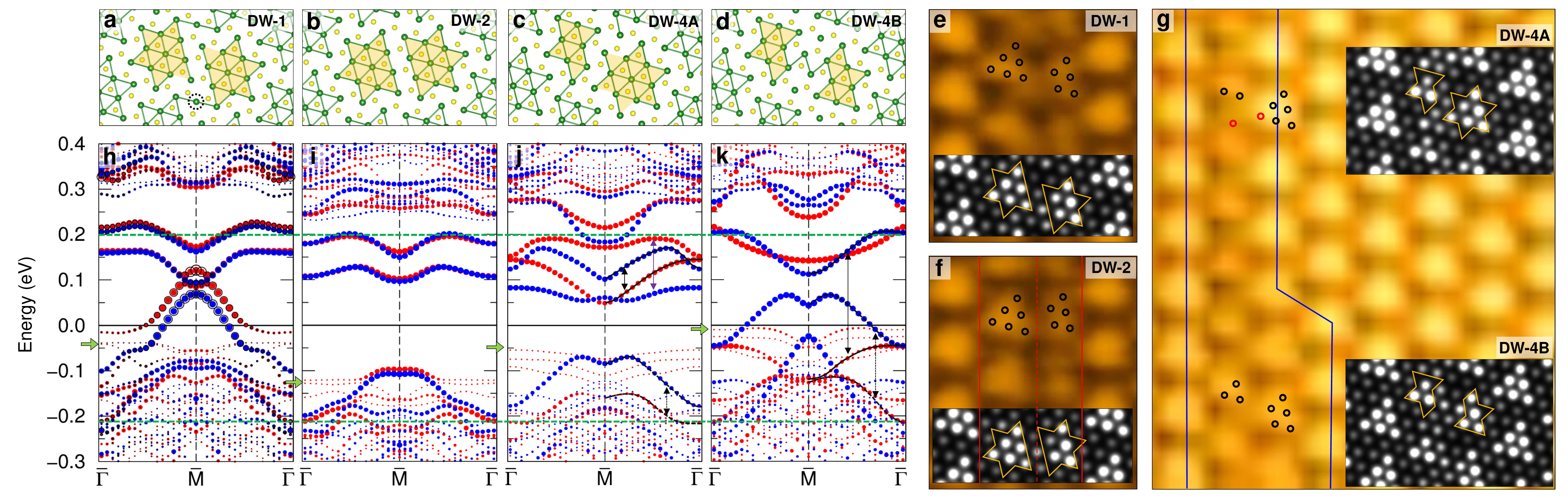} }
\caption{ \label{fig3}
Atomic and electronic structures of major domain walls.
{\bfseries\sffamily a}-{\bfseries\sffamily d} Atomic structure:
{\bfseries\sffamily a} DW-1, {\bfseries\sffamily b} DW-2, {\bfseries\sffamily c} DW-4A, and {\bfseries\sffamily b} DW-4B.
The imperfect DS clusters are marked by the yellow shadows.
{\bfseries\sffamily e}-{\bfseries\sffamily g} Experimental STM image. The circles denote the top S atomic position in atomic structure.
Insets are the theoretical STM simulations.
Simulated STM images are obtained by integrating charge density from -1.2 to 0 eV (Fermi level) of each optimized structure.
The images represent the charge density at a constant height of 2 {\AA} away from the top S atom.
Red lines in {\bfseries\sffamily f} denote the asymmetric double rows.
Blue lines in {\bfseries\sffamily g} denote the domain edge of single (DW-4A) and double rows (DW-4B) domain walls.
{\bfseries\sffamily h}-{\bfseries\sffamily k} Band structures.
Red (Blue) circles denote the spin-majority (minority) states.
The circle size is proportional to the localized states at domain-wall region (shadow DS clusters).
Black open circles in {\bfseries\sffamily h} are localized states at the linked atom marked by the dashed circle in {\bfseries\sffamily a}.
Green arrows indicate the lower Hubbard states at domain region. Green dashed lines indicate the upper and lower Hubbard states of the CCDW phase for comparison.
Purple arrow in {\bfseries\sffamily k} indicate the upper and lower Hubbard states of the domain edge.
Black solid and dashed arrows in {\bfseries\sffamily j} and {\bfseries\sffamily k} are guide for eyes for the spin-polarized bands.
}
\end{figure*}

\subsection*{Electronic structures of straight domain walls.}
The rich variations in DW structures result in a variety of localized electronic states. 
All twelve straight DWs exhibit in-gap states localized one-dimensionally along the DWs (Supplementary Fig. 7).
The Mott gap of the CCDW phase is, thus, substantially reduced or closed at DWs.
However, details of in-gap states depend on various different parameters such as the number of Ta atoms (\textit{d$_z$} electrons) in the DW region, the structural relaxation, the correlation effect, and the charge transfer from neighboring domains.
For example, the DW-2 structure exhibits a rather large band gap (Fig. 3i), which is consistent with the previous STS study \cite{cho17}.
There exist two in-gap states localized along this DW at about 0.1 and 0.2 eV above the Fermi energy.
The majority and minority spin bands for these in-gap states (Fig. 3i) are almost degenerated and their band dispersions do not change substantially with the inclusion of the Coulomb energy U (Supplementary Fig. 7). 
That is, the two in-gap states are split trivially by the bonding-antibonding interaction due to the structural relaxation. 
The insulating property of DW-2 is basically due to the even number of electrons from the two identical imperfect DS unitcells within the DW.

In stark contrast, DW-1 with a similar atomic structure exhibits a metallic behavior with two in-gap bands crossing the Fermi level. 
These bands are due to a unpaired electron on the additional link Ta atom (marked by the black circle in Fig. 3a), which makes the total electron number within DW-1 odd.
As shown in Fig. 3h, the band splitting is due to the local spin and the electron correlation. 
Except for this Ta atom with partially filled spin-polarized states, the atomic configuration with two identical imperfect DSs, each with twelve Ta atoms, in the DW-1 structure is similar to the DW-2 structure and the band dispersions of corresponding in-gap states above the Fermi level are comparable to each other.
This comparison manifests the crucial importance of the number of electrons within a DW unitcell.  


\begin{figure*}[t] 
\centering{ \includegraphics[width=16.0 cm]{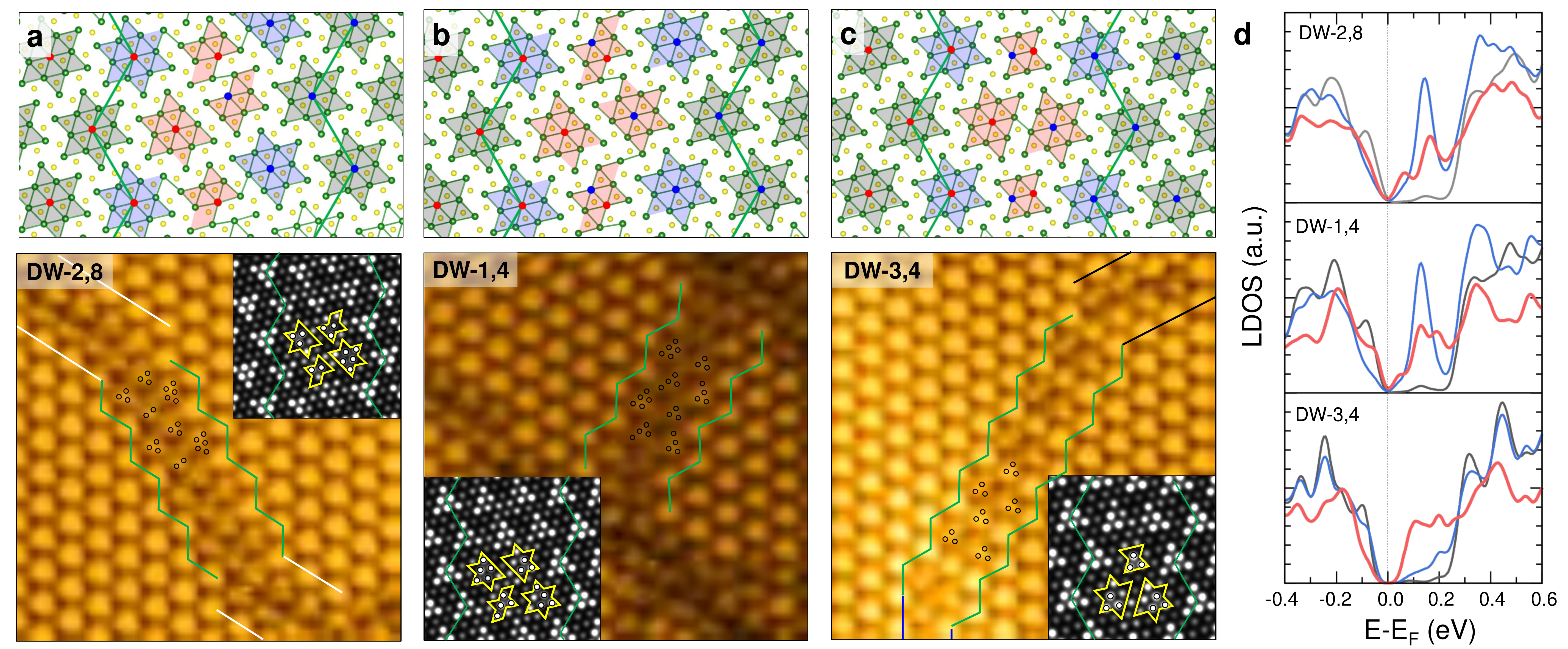} }
\caption{ \label{fig4}
Atomic and electronic structure of $zigzag$ domain walls.
{\bfseries\sffamily a} DW-2,8, {\bfseries\sffamily b} DW-1,4, and {\bfseries\sffamily c} DW-3,4.
Red and blue dots in atomic structures denote the center site of the David-star cluster of left and right domains, respectively.
Bottom panels are experimental STM images.
Insets are simulated STM images as same with Fig. 3. Green zigzag lines denote the domain edge.
{\bfseries\sffamily d} Local DOS of DW-2,8, DW-1,4, and DW-3,4.
Line colors in {\bfseries\sffamily d} correspond to the localized states at the clusters with same color in atomic structures.
}
\end{figure*} 

The atomic and electronic property of a DW can substantially be affected by the charge transfer in addition to the electron correlation.
The single row structure of DW-4A is endowed with an odd number of Ta atoms and the partially filled orbital is expected on the imperfect DS cluster within the DW.
The DW-4A, however, exhibits an insulating character with a band gap of 0.1 eV due to the charge transfer from the edge of neighboring domain to the DW: We find that the fully filled state between -0.2 to -0.1 eV (the blue band) is spin-split and localized on one edge atom of the broken DS cluster of the DW (Supplementary Fig. 8). 
On the other hand, the lower and upper Hubbard states of the domain-edge DS cluster is shifted upward just above the Fermi level (indicated by purple arrow in Fig. 3j), which is consistent with the previous STS spectra \cite{cho17}. These behaviors indicate that electrons are transferred from the edge of the neighboring domains to the DW clusters. 
We further notice that the amount of charge transfer depends on the structural distortion of the DS clusters on domain edges (Supplementary Fig. 9).
In the double row structure (DW-4B) of the same domain wall, the DW width become large with two broken DS clusters. 
A band localized at those imperfect clusters becomes partially filled due to the charge transfer from the domain region (indicated by the green arrow in Fig. 3k). 
The other states on the imperfect DS clusters within the DW open a relatively large band gap of 0.25 eV with spin-split bands (Supplementary Fig. 8), therefore there is a possibility of the 1D magnetism along this DW (Supplementary Fig. 10). However, this is closely related to the controversial issue of the magnetic ordering for the CCDW phase \cite{faze80,law17,klan17,kart17,pal20} thus the magnetic ordering predicted in the present calculations for the DWs has to be investigated further by experiments and more sophisticated calculations.
The above comparison of two different DW structures for one DW type indicates the importance of the charge transfer between DWs and neighboring domains as well as the structural degree of freedom. 
In most of DWs except the DW-11, the Hubbard bands of neighboring domains are shifted to a higher energy indicating the donation of electrons from domains to DW (Table I and Supplementary Fig. 7).
This is consistent with the peak shift of Hubbard states at domain edges observed in the previous experiment \cite{cho17}.

\subsection*{Zigzag domain walls.} 
Figure 4 shows the atomic structures and STM images of $zigzag$ DWs (DW-2,8, DW-1,4, and DW-3,4).
The DW-2,8 and DW-1,4 consist of four types of Ta clusters and the clusters are connected to each other though distortions of outer Ta atoms.
The DW-3,4 consists of three types of Ta clusters so that the width is narrower.
These STM topographies, which are well reproduced in our DFT simulations, reveal that the $zigzag$ DWs have an even larger structural degree of freedom.
There exist two types of in-gap states related to these $zigzag$ DWs, which are localized at DWs (the red lines in Fig. 4d obtained on the broken DW clusters mentioned above, which is indicated by red clusters in Fig. 4a, b and c) and edges of neighboring domains (blue lines of Fig. 4d and blue clusters in Fig. 4a, b and c), respectively.
These states cover most of the Mott gap but the local density of states (LDOS) of the whole systems commonly touch zero at the Fermi level, which is close to a pseudogap. 
Note that the $zigzag$ DWs has the $\times$2-ordered Ta clusters along the DW direction.
We attribute the common insulating nature of $zigzag$ DWs to the alternating hopping amplitudes similar to the insulating 1D $zigzag$ dimer chain of the Su-Schrieffer-Heeger model \cite{su79}, while we do not establish a quantitative model.
The notable difference between DW-3,4 and DW-1,4/DW-2,8 is that the domain edge states of DW-3,4 preserve the Mott insulating character with upper and lower Hubbard states largely while those on DW-2,8 and DW-1,4 show a large in-gap states at around 0.1 eV without strong spin polarization (Supplementary Fig. 11).
This is related to how the DS clusters on domain edges are distorted due to the neighboring imperfect DS clusters within DWs. 

\section*{Discussion}

This work provides a totally different approach and understanding from a recent theoretical work, which explained the insulating properties of $straight$ DWs of 1T-TaS$_2$ in terms of the frustrated hopping within the triangular Mott insulator model \cite{skol19}. 
In fact, the particular type of DWs mentioned in the previous model is metallic in our experiment and calculation. 
This discrepancy indicates clearly the limitation of this simple model, which ignores all the internal degrees of freedom of DWs.

This work uses the large tunneling current and bias to induce DW networks. 
While the mechanism is not fully clear yet, the network may be induced by the local heating or the electric field. 
The DW phases were also induced globally by different methods of by heating \cite{park19} and optical pulses \cite{stoj14}. 
Since each method may have different levels of excitation and thermodynamics, the lateral and population distribution of different types of DWs induced can be different.
For example, a recent STM study showed that the optical pulse produces irregular networks of the $zigzag$ DWs and DW-2 configurations \cite{gera19}.
This is partly consistent with our work.
On the other hand, as we reported recently, the thermally excited DW phase (the NCCDW phase above 180 K) has only the DW-1 type uniformly \cite{park19}.
Nevertheless, what is very important here is that we tabulate most of the DW structures possible and provide a unified understanding of their properties.

We investigate systematically structural and electronic properties of DW structures of the 1T-TaS$_{2}$ surface layer, which was manipulated by scanning tunneling microscope (STM) voltage pulses to generate DW networks with a plethora of metastable DWs \cite{cho16,ma16}. 
We identify more than ten different DW structures including the straight DWs discussed partly in the previous works, the newly observed $zigzag$ DWs, and the breakdown of an unstable DWs into multiple DWs of a lower energy. 
The atomic structures and electronic states of most of these DWs are determined by direct comparisons of STM data and density functional theory (DFT) calculations including the on-site Coulomb interaction. 
Various types of DW configurations lead to rich electronic states in narrow energy scale within the band gap of 1T-TaS$_{2}$. 
A systematic understanding of the structural and electronic variety of these DWs are obtained through extensive DFT calculations.
The electronic property of these discommensurate DWs exhibits both metallic and insulating characters depending on their atomic structure, correlation, and charge transfer from neighboring domains.
Our finding of the DW properties would help to understand microscopic origins of various metastable phases, hidden phases, and emerging superconductivity, which are all attributed to the electronic states localized on DWs. 

\section*{Methods}

\subsection*{Sample preparations}
The single crystal 1T-TaS$_{2}$ were grown by iodine vapour transport method in the evacuated quartz tube.
Before growth of the sample, the powder 1T-TaS$_{2}$ was sintered for 48 h at 750 $^{\circ}$C.
We repeated this process two times to get poly crystals.
To get high-quality sample, the seeds were slowly transported by iodine at 900$\sim$970 $^{\circ}$C for 2 weeks.
The tube was rapidly cooled down to room temperature in the air due to the metastability of the 1T phase.

\subsection*{Experiments}
The STM measurements were carried out at $T$ = 4.3 K with a commercial STM (SPECS) and the mechanically sharpened Pt–Ir wires were used for STM tips.
All of STM images were obtained with the constant current mode and the bias voltage $V_{s}$ was applied to the sample.
The detailed method of the textured CDW domain formation and the preparation of single crystal 1T-TaS$_{2}$ are similar to previous study \cite{cho16}.

\subsection*{DFT calculations}
The DFT calculations were carried out by using the Vienna $ab$ $initio$ simulation package \cite{kres96} within the Perdew-Burke-Ernzerhof functional of the generalized gradient approximation \cite{perd96} and the projector augmented wave method \cite{bloc94}.
We used the single-layer model of 1T-TaS$_{2}$ with a vacuum spacing of about 13.6 {\AA}.
We assumed the 2D-like Mott states at the surface limit based on most of DWs are not coupled between the first and the second layer (Supplementary Fig. 12) and the interlayer coupling of the DWs in a bilayer structure is not affecting substantially the electronic structure of the DWs (Supplementary Fig. 13).
A plane-wave basis set of 259 eV and a 6$\times$6$\times$1 $k$-point mesh for the  $\sqrt{13}\times\sqrt{13}$ unit cell are used.
All atoms were relaxed until the residual force components became smaller than 0.02 eV/{\AA}.
An on-site Coulomb energy (U = 2.3 eV) was included for Ta 5$d$ orbitals to reproduce the Mott gap size in experiment \cite{cho15}.
Similar calculation methods were successfully used in our previous study \cite{park19}.


\section*{Acknowledgements}
This work was supported by Institute for Basic Science (Grant No. IBS-R014-D1).

\section*{Author contributions}
H.W.Y. supervised the research. J.W.P. performed the DFT calculations. J. L. carried out the STM experiment. J.W.P and H.W.Y. analyzed the data, and wrote the manuscript with the comments of all other authors.

\section*{Competing interests}
The Authors declare no Competing interests.


\newcommand{\AP}[3]{Adv.\ Phys.\ {\bf #1}, #2 (#3)}
\newcommand{\CMS}[3]{Comput.\ Mater.\ Sci. \ {\bf #1}, #2 (#3)}
\newcommand{\PR}[3]{Phys.\ Rev.\ {\bf #1}, #2 (#3)}
\newcommand{\PRL}[3]{Phys.\ Rev.\ Lett.\ {\bf #1}, #2 (#3)}
\newcommand{\PRB}[3]{Phys.\ Rev.\ B\ {\bf #1}, #2 (#3)}
\newcommand{\NA}[3]{Nature\ {\bf #1}, #2 (#3)}
\newcommand{\NAP}[3]{Nat.\ Phys.\ {\bf #1}, #2 (#3)}
\newcommand{\NAM}[3]{Nat.\ Mater.\ {\bf #1}, #2 (#3)}
\newcommand{\NAC}[3]{Nat.\ Commun.\ {\bf #1}, #2 (#3)}
\newcommand{\NAN}[3]{Nat.\ Nanotechnol.\ {\bf #1}, #2 (#3)}
\newcommand{\NARM}[3]{Nat.\ Rev.\ Mater.\ {\bf #1}, #2 (#3)}
\newcommand{\NL}[3]{Nano \ Lett.\ {\bf #1}, #2 (#3)}
\newcommand{\NT}[3]{Nanotechnology {\bf #1}, #2 (#3)}
\newcommand{\JP}[3]{J.\ Phys.\ {\bf #1}, #2 (#3)}
\newcommand{\JAP}[3]{J.\ Appl.\ Phys.\ {\bf #1}, #2 (#3)}
\newcommand{\JPSJ}[3]{J.\ Phys.\ Soc.\ Jpn.\ {\bf #1}, #2 (#3)}
\newcommand{\PNAS}[3]{Proc.\ Natl.\ Acad.\ Sci. {\bf #1}, #2 (#3)}
\newcommand{\PRSL}[3]{Proc.\ R.\ Soc.\ Lond. A {\bf #1}, #2 (#3)}
\newcommand{\PBC}[3]{Physica\ B+C\ {\bf #1}, #2 (#3)}
\newcommand{\PAC}[3]{Pure Appl.\ Chem. \ {\bf #1}, #2 (#3)}
\newcommand{\SCI}[3]{Science\ {\bf #1}, #2 (#3)}
\newcommand{\SCA}[3]{Sci.\  Adv.\ {\bf #1}, #2 (#3)}
\newcommand{\RPP}[3]{Rep.\ Prog.\ Phys. \ {\bf #1}, #2 (#3)}
\newcommand{\NPJQ}[3]{npj \ Quantum \ Materials \ {\bf #1}, #2 (#3)}


\begin{thebibliography}{}
\bibitem{wood14} C. R. Woods, L. Britnell, A. Eckmann, R. S. Ma, J. C. Lu, H. M. Guo, X. Lin, G. L. Yu, Y. Cao, R. V. Gorbachev, A. V. Kretinin, J. Park, L. A. Ponomarenko, M. I. Katsnelson, Yu. N. Gornostyrev, K. Watanabe, T. Taniguchi, C. Casiraghi, H-J. Gao, A. K. Geim and K. S. Novoselov, Commensurate-incommensurate transition in graphene on hexagonal boron nitride, \NAP{10}{451}{2014}.
\bibitem{cao18} Y. Cao, V. Fatemi, S. Fang, K. Watanabe, T. Taniguchi, E. Kaxiras, and P. Jarillo-Herrero, Unconventional superconductivity in magic-angle graphene superlattices, \NA{556}{43}{2018}.
\bibitem{naik2018} M. H. Naik and M. Jain, Ultraflatbands and shear solitons in Moir{\'e} patterns of twisted bilayer transition metal dichalcogenides, \PRL{121}{266401}{2018}.
\bibitem{faze80} P. Fazekas and E. Tosatti, Charge carrier localization in pure and doped 1T-TaS$_2$, \PBC{99}{183}{1980}.
\bibitem{may11} M. M. May, C. Brabetz, C. Janowitz, and P. Manzke, Charge-density-wave phase 1T-TiSe$_2$: The influence of conduction band population, \PRL{81}{176405}{2011}.
\bibitem{yan17} S. Y. Yan, D. Iaia, E. Morosan, E. Fradkin, P. Abbamonte, and V. Madhavan, Influence of domain walls in the incommensurate charge density wave state of Cu intercalated 1T-TiSe$_2$ \PRL{118}{106405}{2017}.
\bibitem{joe14} Y. I. Joe, $et$ $al.$, Emergence of charge density wave domain walls above the superconducting dome in 1T-TiSe$_2$ \NAP{10}{2935}{2014}.
\bibitem{xu10} P. Xu, J. O. Piatek, P.-H. Lin, B. Sipos, H. Berger, L. Forr{\'o}, H. M. R$\phi$nnow, and M. Grioni, Superconducting phase in the layered dichalcogenide 1T-TaS$_2$ upon inhibition of the metal-insulator transition, \PRB{81}{172503}{2010}.
\bibitem{Ang12} R. Ang, R. Ang, Y. Miyata, E. Ieki, K. Nakayama, T. Sato, Y. Liu, W. J. Lu, Y. P. Sun, and T. Takahashi Superconductivity and bandwidth-controlled Mott metal–insulator transition in 1T-TaS$_{2-x}$Se$_x$, \PRB{88}{115145}{2013}.
\bibitem{stoj14} L. Stojchevska, I. Vaskivskyi, T. Mertelj, P. Kusar, D. Svetin, S. Brazovskii, D. Mihailovic, Ultrafast switching to a stable hidden quantum state in an electronic crystal, \SCI{344}{177}{2014}.
\bibitem{vask15} I. Vaskivskyi, J. Gospodaric, S. Brazovskii, D. Svetin, P. Sutar, E. Goreshnik, I. A. Mihailovic, T. Mertelj, D. Mihaiovic, Controlling the metal-to-insulator relaxation of the metastable hidden quantum state in 1T-TaS$2$, \SCA{1}{e1500168}{2015}.
\bibitem{vask16} I. Vaskivskyi,  I. A. Mihailovic, S. Brazovskii, J. Gospodaric, T. Mertelj, D. Svetin, P. Sutar, and D. Mihailovic, Fast electronic resistance switching involving hidden charge density wave states, \NAC{7}{11442}{2016}.
\bibitem{han15} T.-R. T. Han, F. Zhou, C. D. Malliakas, P. M. Duxbury, S. D. Mahanti, M. G. Kanatzidis, and C.-Y. Ruan, Exploration of metastability and hidden phases in correlated electron crystals visualized by femtosecond optical doping and electron crystallography, \SCA{1}{e1400173}{2015}.
\bibitem{yosh15} M. Yoshida, R. Suzuki, Y. Zhang, M. Nkano, Y. Iwasa, Memristive phase switching in two-dimensional 1T-TaS$_2$ crystals, \SCA{1}{e1500606}{2015}.
\bibitem{tsen15} A. W. Tsen, $et$ $al.$, Structure and control of charge density waves in two-dimensional 1T-TaS$_2$, \PNAS{112}{15054}{2015}.
\bibitem{zong18} A. Zong, $et$ $al.$, Ultrafast manipulation of mirror domain walls in a charge density wave, \SCA{4}{eaau5501}{2018}
\bibitem{wen19}W. Wen, C. Dang, and L. Xie, Photoinduced phase transitions in two-dimensional charge-density-wave 1T-TaS$_{2}$, Chin. Phys. B {\bf 28}, 058054 (2019).
\bibitem{cho16} D. Cho,  S. Cheon, K.-S. Kim, S.-H. Lee, Y.-H. Cho, S.-W. Cheong, and H. W. Yeom, Nanoscale manipulation of the Mott insulating state coupled to charge order in 1T-TaS$_2$, \NAC{7}{10453}{2016}.
\bibitem{ma16} L. Ma, C. Ye, Y. Yu, X. F. Lu, X. Niu, S. Kim, D. Feng, D Tom{\'a}nek, Y.-W. Son, X, H, Chen, and Y. Zhang, A metallic mosaic phase and the origin of Mott insulating state in 1T-TaS$_2$. \NAC{7}{10956}{2016}.
\bibitem{bu19} K. Bu, W. Zhang, Y. Fei, Z. Wu, Y. Zheng, J. Gao, X. Luo, Y.-P. Sun, and Y. Yin, Possible strain induced Mott gap collapse in 1T-TaS$_{2}$, Comms. Phys. {\bf 2}, 146 (2019).
\bibitem{saka17} D. Sakabe, Z. Liu, K. Suenaga, K. Nakatsugawa, and S. Tanda, Direct observation of mono-layer, bi-layer, and tri-layer charge density waves in 1T-TaS$_2$ by transmission electron microscopy without a substrate, \NPJQ{2}{22}{2017}.
\bibitem{laul17} C. Laulh{\' e}, $et$ $al.$, Ultrafast formation of a charge density wave state in 1T-TaS$_2$: Observation at nanometer scales using time-resolved X-ray diffraction, \PRL{118}{247401}{2017}.
\bibitem{avig18} I. Avigo, P. Zhou, M. Kall{\"a}ne, K. Rossnagel, U. Bovensiepen, and M. Ligges, Excitation and relaxation dynamics of the photo-perturbed correlated electron system 1T-TaS$_2$, Appl. Sci. {\bf 9}, 44 (2019).
\bibitem{ang12} R. Ang, Y. Tanaka, E. Ieki, K. Nakayama, T. Sato, L. J. Li, W. J. Lu, Y. P. Sun, and T. Takahashi, Real-space coexistence of the melted Mott state and superconductivity in Fe-substituted 1T-TaS$_2$, \PRL{109}{176403}{2012}.
\bibitem{liu16} Y. Liu, $et$ $al.$,Nature of charge density waves and superconductivity 1T-TaSe$_{2-x}$Te$_{x}$, \PRB{94}{045131}{2016}.
\bibitem{sipo08} B. Sipos, A. F. Kusmartseva, A. Akrap, H. Berger, L. Forr{\'o}, and E. Tutis{\'s}, From Mott state to superconductivity in 1T-TaS$_2$, \NAM{7}{960}{2008}.
\bibitem{li16} L. J. Li, E. C. T. O’Farrell, K. P. Loh, G. Eda, B. {\"O}zyilmaz, and A. H. Castro Neto, Controlling many-body states by the electric-field effect in a two-dimensional material \NA{529}{185}{2016}.
\bibitem{chen18} C. Chen, L. Su, A. H. Castro Neto, Vitor M. Pereira, Discommensuration-enhanced superconductivity in the charge density wave phases of transition-metal dichalcogenides, \PRB{99}{121108(R)}{2019}.
\bibitem{park19} J. W. Park, G. Y. Cho, J. Lee, and H. W. Yeom, Emergent honeycomb network of topological excitations correlated charge density wave \NAC{10}{4038}{2019}.
\bibitem{pill00} Th. Pillo, J. Hayoz, H. Berger, R. Fasel, L. Schlapbach, and P. Aebi, Interplay between electron–electron interaction and electron–phonon coupling near the Fermi surface of 1T-TaS$_2$, \PRB{62}{4277}{2000}.
\bibitem{cler06} F. Clerc, C. Battaglia, M. Bovet, L. Despont, C. Monney, H. Cercellier, M. G. Garnier, P. Aebi, H. Berger, and L. Forr{\'o}, Lattice-distortion-enhanced electron-phonon coupling and Fermi surface nesting in 1T-TaS$_2$, \PRB{74}{155114}{2006}.
\bibitem{cho15} D. Cho, Y.-H. Cho, S.-W. Cheong, K.-S. Kim, and H. W. Yeom, Interplay of electron-electron and electron-phonon interactions in the low-temperature phase of 1T-TaS$_2$, \PRB{92}{085132}{2015}.
\bibitem{law17} K. T. Lawa and P. A. Lee, 1T-TaS$_2$ as a quantum spin liquid, \PNAS{114}{6996}{2017}.
\bibitem{klan17} M. Klanj{\v s}ek, A. Zorko, R. {\v Z}itko, J. Mravlje, Z. Jaglicic, P. K. Biswas, P. Prelov{\v s}ek1,5, D. Mihailovic and D. Arcon, A high-temperature quantum spin liquid with polaron spins, \NAP{13}{1130}{2017}.
\bibitem{brou80} R. Brouwer and F. Jellinek, The low-temperature superstructures of 1T-TaSe$_2$ and 2H-TaSe$_2$, \PBC{99}{51}{1980}.
\bibitem{cho17} D. Cho, G. Gye, J. Lee, S.-H. Lee, L. Wang, S.-W. Cheong, and H. W. Yeom, Correlated electronic states at domain walls of a Mott-charge-density-wave insulator 1T-TaS$_2$, \NAC{8}{392}{2017}.
\bibitem{wu89} X. L. Wu and C. M. Lieber, Hexagonal domain-like charge density wave phase of TaS$_2$ determined by scanning tunneling microscopy, \SCI{243}{1704}{1989}.
\bibitem{naka77} K.Nakanishi, H. Takatera, Y. Yamada, and H. Shiba, The nearly commensurate phase and effect of harmonics on the successive phase transition in 1T-TaS$_2$, \JPSJ{43}{1509}{1977}.
\bibitem{yama83} A. Yamamoto, Hexagonal domainlike structure in 1T-TaS$_2$, \PRB{27}{7823}{1983}.
\bibitem{wu90} X. L. Wu and C. M. Lieber, Direct observation of growth and melting of the hexagonal-domain charge-density-wave phase in 1 T-TaS$_2$ by scanning tunneling microscopy, \PRL{64}{1150}{1990}.
\bibitem{burk91} B. Burk, R. E. Thomson, A. Zettl, and J. Clarke, Charge-density-wave domains in 1T-TaS$_2$ observed by satellite structure in scanning- tunneling-microscopy images, \PRL{66}{3040}{1991}.
\bibitem{spij97} A. Spijkerman, J. L. de Boer, A. Meetsma, and G. A. Wiegers, X-ray crystal-structure refinement of the nearly commensurate phase of 1T-TaS$_2$ in (3+2)-dimensional superspace, \PRB{56}{1357}{1997}.
\bibitem{lee20} J. M. Lee, C. Geng, J. W. Park, M. Oshikawa, S.-S. Lee, H. W. Yeom, and G. Y. Cho, Stable flatbands, topology, and superconductivity of magic honeycomb networks, \PRL{124}{137002}{2020}.
\bibitem{kart17} M. Kratochvilova, A. D. Hillier, A. R. Wildes, L. Wang, S.-W. Cheong, and J.-G. Park, The low-temperature highly correlated quantum phase in the charge-density-wave 1T-TaS$_{2}$ compound, \NPJQ{2}{42}{2017}.
\bibitem{pal20} S. Pal, K. Kumar, R. Sharma, A. Banerjee, S. B. Roy, J.-G. Park, A. K. Nigam and S.-W. Cheong, Possible glass-like random singlet magnetic state in 1T-TaS$_{2}$, J. Phys.: Condens. Matter {\bf 32}, 035601 (2020).
\bibitem{su79} W. P. Su, J. R. Schrieffer, and A. K. Heeger, Solitons in plyacetylene, \PRL{42}{1698}{1979}.
\bibitem{skol19}  J. Skolimowski, Y. Gerasimenko, and R. {\v Z}itko, Mottness collapse without metallization in the domain wall of the triangular-lattice Mott insulator 1T-TaS$_{2}$, \PRL{122}{036802}{2019}.
\bibitem{gera19} Y, A. Gerasimenko, P. Karpov, I. Vaskivskyi, S. Brazovskii, and D. Mihailovic, Intertwined chiral charge orders and topological stabilization of the light-induced state of a prototypical transition metal dichalcogenide, \NPJQ{4}{32}{2019}.
\bibitem{kres96} G. Kresse and J. Furthm{\"u}ller, Efficient iterative schemes for $ab$ $initio$ total-energy calculations using a plane-wave basis set, \PRB{54}{11169}{1996}.
\bibitem{perd96} J. P. Perdew, K. Burke, and M. Ernzerhof, Gerneralized gradient approximation made simple, \PRL{77}{3865}{1996}.
\bibitem{bloc94} P. E. Bl{\"o}chl, Projector augmented-wave method, \PRB{50}{17953}{1994}.

\end{thebibliography}
\end{document}